\documentclass[aps,prl,twocolumn,showpacs,floatfix]{revtex4-1}
\usepackage{graphicx}

\begin{document}

\title{Critical phenomena in the bifurcation line's perspective}

\author{A. Kashuba}

\affiliation{Bogolyubov Institute for Theoretical Physics, 14b Metrolohichna, 
Kiev 03680 Ukraine}

\date{\today}

\begin{abstract}
On the phase diagram of a system undergoing a continuous phase transition of the second order, three lines, hyper-surfaces, convergent into the critical point feature prominently: the  ordered and disordered phases in the thermodynamic limit, and a third line, extending into a domain of finite-size systems, defined by the bifurcation of the distribution of the order parameter. Unlike critical phenomena in the thermodynamic limit devoid of known thermodynamic potential and described rather by the conformal symmetry, in finite-size systems near the bifurcation line an explicit Hamiltonian for the zero-mode of the order parameter is found. It bears the impress of the universality class of the critical point in terms of the two critical exponents: $\beta$ and $\nu$.
\end{abstract}

\pacs{05.70.Jk, 64.60.-i} 

\maketitle

Phase transitions of the second order fascinate us not least because of peculiar critical phenomena, a set of power law relationships, all reaching singularity at the critical point, between thermodynamic observables, like order parameter, and thermodynamic variables, like temperature \cite{dg}. Remarkably, critical exponents in these power law relationships are the same, universal, for many apparently different systems like magnets and vapor-liquid mixtures. Historically, the first, Landau's, theory of phase transitions of the second order was based on an explicit ansatz for the Gibbs thermodynamic potential and the mean-field critical phenomena was predicted \cite{ll5}. In many experiments over years, though, quite different critical phenomena have been observed \cite{dg,l96}. The crucial role of Orenstein-Zernicke thermodynamic fluctuations of the order parameter has been recognized as responsible for this discrepancy \cite{w71}. Exactly at the critical point the fluctuations are so strong that, in principle, there is no predictability for outcome of a measurement even in the thermodynamic limit of infinitely large system. There is no known effective, the so-called coarse-grained, thermodynamic potential to describe the probability distribution for the critical fluctuations, reflecting, in part, the difficulty of satisfying many singular scaling relationships \cite{k66}. Rather, it is known that the critical fluctuations do obey the conformal symmetry \cite{p70}. In two dimensions the representations theory of the conformal symmetry can provide values for the critical exponents \cite{conf}.

Finite-size systems are more promising for a theory based on the effective thermodynamic potential. First, in finite-size systems there is no critical phenomena as such and no singularities in particular. Second, a distribution of any fluctuating observable for a whole system is a smooth function and, if known, gives an effective potential directly. The finite-size scaling theory \cite{p} predicts that one dependency variable, for instance the size of the system $L$, is redundant and observables depend on the dimensionless ratio of $L$ to the correlation length $\xi(T)$ instead. The distribution of the order parameter has been proposed to be universal at exactly the thermodynamic critical temperature $T_c$  \cite{b81}. However, the anisotropy of the sample shapes \cite{ff69} or anisotropy of atomic bond interactions \cite{ss09} render it non-universal. For Ising models, the distribution of magnetization is related to a partition function in imaginary external magnetic field. In this way the exact distribution of magnetization at $T_c$ has been found in the ground state of the one dimensional quantum Ising chain in transverse field \cite{lf08}.

Where the phase transition of the second order is centered on in finite-size systems: either on the thermodynamic critical temperature $T_c$ or on a line where some divergent observable, like the specific heat, reaches a maximum \cite{ff69}? In the spirit of the Landau theory \cite{ll5} it is a temperature $T_c(L)$ where the distribution of the order parameter transforms, when the system is being cooled down, from the maximum to the minimum at zero order parameter. This phenomenon, called a bifurcation, if viewed on short time scales, characteristic for the kinetics of small systems, does represent a change of the symmetry of the system's current thermal state, like that in the Landau theory in the thermodynamic limit \cite{ll5}.

In this letter the effective potential, the distribution of the order parameter, is found on and around the bifurcation line. It dependence on the order parameter is new one and exact one. The effective potential explicitly displays $\beta$, the critical exponent of magnetization, and faithfully represents the universality class of the adjacent thermodynamic critical point. 

The physical properties of small clusters are important in many applications of nano-technologies. Magnetic clusters may show a correlated behavior as a one effective spin near the phase transition of the second order. The total magnetization response of a critical cluster to a uniform magnetic field is a critical phenomenon depending on the temperature and the field strength. Initially, at small fields a cluster is responding as a whole according to the linear law. As the magnetic field exceeds some crossover field and the correlation length shrinks to be within the cluster size, the response becomes local and follows a non-linear power law: $m\propto h^{1/\delta}$, specified by the critical exponent of the field $\delta$ \cite{ll5}. This letter predicts that an enhancement over this response develops with the peak being reached on the bifurcation line: $m_c\propto h^{\beta/(1+\beta)}$.

In a uniform system described by a Hamiltonian $H[\sigma]$, the functional of microscopic states $\{\sigma\}$ completely describing the system, the probability to find a state $\{\sigma\}$ at some temperature $T$ is given, in statistical physics, by a weight $w=\exp (-H[\sigma]/T)/Z$, where $Z$ is the corresponding partition function. If one introduces a local projection of the state $\{\sigma\}$ onto the atomic level at site $\mathbf{x}$ as $\sigma_\mathbf{x}$, then correlation functions like the two-point one: $\langle \sigma_\mathbf{x} \sigma_\mathbf{y} \rangle$, can be defined. Now, we cut out a cluster of finite size $L$ from this system. Neglecting the effect of the special profile of the order parameter at interfaces, we introduce the total order parameter: $M= \sum_\mathbf{x} \sigma_\mathbf{x}$, the effective zero-dimensional mode of the cluster, with a probability distribution function being defined as:
\begin{equation}
P(M)=\frac{1}{Z}\sum_{\{\sigma\}} \delta\left( \sum_\mathbf{x} \sigma_\mathbf{x} -M \right) e^{-H[\sigma]/T} .
\end{equation}
It is normalized: $\sum_M P(M)=1$. Interchangeably, we will use the distribution of the order parameter per site $m=M/L^d$: $P(m)$. The distribution function of the order parameter serves for small cluster as a probability weights of the effective zero-mode $m$ with the effective potential being defined as:
\begin{equation}\label{ZM}
H(m,T)=-T\log P(m)
\end{equation}
Unlike microscopic Hamiltonian it depends on the temperature. For the zero-mode, correlation functions, moments of the order parameter: $\langle |m|^n \rangle$, can be defined and studied too. Statistical average of the free energy: $\langle H(m,T) \rangle$ allows one to construct the thermodynamics of the finite-size cluster in the usual way \cite{ll5}.

\begin{figure} \includegraphics[width=0.4\textwidth]{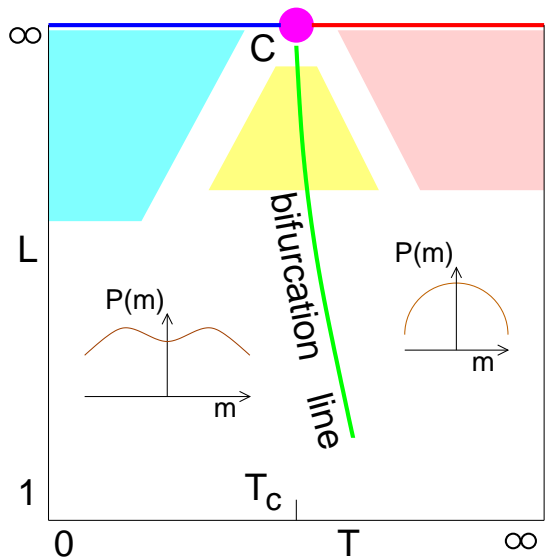}
\caption{Schematic phase diagram of the critical system in the plane of temperature and finite-size: $(T,L)$. (Color online)}
\label{PhaseDiagram} \end{figure}

Consider a typical phase diagram of a system undergoing a phase transition of the second order at some critical point $C$ in the thermodynamic limit and extend it into a domain of finite clusters of the same system with varying sizes, see Fig.\ref{PhaseDiagram}. For definiteness we will discuss and will do numerical simulations for the three-dimensional Ising model with nearest neighbor interaction:
\begin{equation}\label{Ham}
H[\sigma]=-J\sum_{\langle \mathbf{x}\mathbf{y}\rangle} \sigma_\mathbf{x} \sigma_\mathbf{y}
\end{equation}
where local spin takes on two values $\sigma_\mathbf{x}=\pm 1$ and $\mathbf{x}$ numerates sites of the three-dimensional, $d=3$, simple cubic lattice. The Ising Hamiltonian is invariant under the global $Z_2$ symmetry transformation. In the finite-size cluster the true phase transition is impossible and in the long-time a thermal state of the cluster will conserve the global $Z_2$ symmetry of the Ising model. Therefore, $P(m)$ is an even function of $m$.

There is no analytic functions $P(m)$ and $H(m,T,L)$ rather a set of values at equidistant points $m$ with no limiting point. Therefore, there is no unique analytic continuation into a complex plane $m$. We can fit this set using a function: $\sum_k a_k(T,L) m^{2k}$, with the number of terms being equal to the number of points $m$. However, such function will swing wildly in between points $m$. By increasing the number of coefficients, the function $\sum_k a_k m^{2k}$ can be made smoother. However, in the complex plane $m$ close to the real axis singularities, like poles and cuts, will develop. Therefore, a power-like function: $H(m,T)\propto |m|^{1+\Delta}$, where $|m|$ is understood symbolically as $\sqrt{m^2+1/L^{2d}}$, could not be ruled out from the first principles in the limit $L\to \infty$.

As a function of temperature, though, $H(m,T)$ is a smooth analytic function. At low temperatures it has two minima at $m=\pm m_0$ and one maximum at $m=0$. At high temperature it has one minimum at $m=0$. At some intermediate temperature $T_c(L)$, depending on the cluster size $L$, the two minima $\pm m_0$ and the maximum $m=0$ will merge into one minimum $m=0$ in what is usually known as a bifurcation phenomenon. The bifurcation line $T_c(L)$ for finite size clusters is a closest analog of the critical point $T_c$ in the thermodynamic limit of infinite system. Since the bifurcation line has the end point in the critical point, see Fig.\ref{PhaseDiagram}, the finite-size scaling relationships holds:
\begin{equation}\label{BifLine}
T_c(L)=T_c+\theta_c/L^{1/\nu} ,
\end{equation}
where $\nu$ is the critical exponent for the correlation length and $T_c$ and $\theta_c$ are two parameters that determine the bifurcation line completely. Both are non-universal. For instance, if one adds next nearest interactions to the Ising Hamiltonian Eq.(\ref{Ham}) then $T_c$ will change. Similarly, if one modifies the boundary conditions or stretches the perfect cubes then $\theta_c$ will change too. 

We derive zero-mode Hamiltonian using two principles: {\it the larger is the size the more severe are singularities; the zero-mode Hamiltonian should display the universality class of the critical point.} The first principle reflects the nature of substantial singularity of the thermodynamic potential in the critical point. Since the yellow sector encompassing the bifurcation line on the phase diagram in Fig.\ref{PhaseDiagram} ends up in the critical point, the second principle asserts that all the properties of the critical point should be revealed by continuation from this sector. In the finite-size system it is proportions of the effective zero-mode Hamiltonian rather than the critical phenomena that give the critical exponents.

When the temperature changes away from the bifurcation line Eq.(\ref{BifLine}) the variation of the zero-mode Hamiltonian, with the Zeeman term being added, is analytic and expandable into the Taylor series in $t=T-T_c(L)$:
\begin{equation}
H(m,T,L)=H_0(m,L)+t H_1(m,L) -hM .
\end{equation}
In the thermodynamic limit: $L\to \infty$, there exists the Widom scaling function $f(x)$ such that
\begin{equation}
h=|m|^\delta f\left( t/|m|^{1/\beta} \right) ,
\end{equation}
with the property: $f(x)=(\delta-\gamma/\beta+1) C^{-1}_+ x^\gamma$ as $x\to \infty$. Integrating this relationship from $m=0$ to some small $m$ at some positive $t$, above $T_c$, and using the Widom's scaling relationship: $\beta\delta=\beta+\gamma$, we find:
\begin{equation}
\Phi(m,T)=\Phi_0(m,T)+ L^d C^{-1}_+ t^\gamma m^2 -hM .
\end{equation}
We see that already in the thermodynamic limit the dependence of this term on the magnetization is analytic. Therefore, invoking the first principle we conclude that the first term of the Taylor expansion of the zero-mode Hamiltonian away from the bifurcation line is analytic too: $H_1(m,L)\propto m^2$. Returning to the thermodynamic limit at exactly the critical temperature $t=0$ we obtain:
\begin{equation}
\Phi_0(m,T_c)=\Phi_0+ L^d  f(0) |m|^{1+\delta}/ (1+\delta) .
\end{equation}
Now, we invoke the second principle. Scaling relationships: Widom's, Rushbrook's $\alpha+ 2\beta+ \gamma=2$, Fisher's $(2-\eta)\nu= \gamma$ and Josephson's $d\nu =2-\alpha$, supply all critical exponents from a minimum of the two critical exponents that must be present in the zero-mode Hamiltonian. The first one is the critical exponent for the correlation length: $\nu$, that controls the finite-size scaling. The second one is related to the new critical exponent $\Delta$ for the distribution of the magnetization:  
\begin{equation} \label{proportion}
H(m,T,L)=B(T,L) |m|^{1+\Delta} +a(T,L) t m^2 -hM .
\end{equation}
We are presented here with a dilemma. The critical point's proportion relates either the first and the third terms or the first and the second terms: $\Delta=\delta$ or $\Delta=1+1/\beta$, but not the both.

\begin{figure} \includegraphics[width=0.48\textwidth]{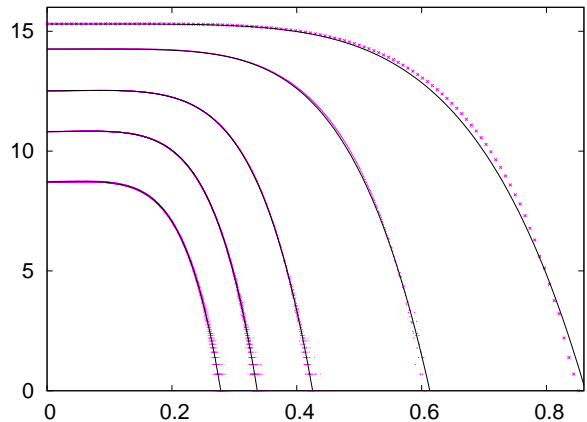}
\caption{Logarithm of the distribution function of magnetization versus the magnetization 
per lattice site on the bifurcation line for five cubes of the three-dimensional Ising model of sizes $8^3$, $16^3$, $32^3$, $48^3$ and $64^3$ in descending order. (Color online)} \label{Phi} \end{figure}

Numerical simulations of the three-dimensional Ising model using the Metropolis update of the state of the system immediately resolve this dilemma in favor of $\Delta=1+1/\beta$. We notice that for studying the distribution of magnetization the critical slowing down is compensated by the time it is needed to fill the histogram. Five clusters with periodic boundary conditions, of sizes $L=8,16,32,48,64$, were studied at temperatures adjusted to be as close as possible to the bifurcation line $T_c(L)$. The results are shown in Fig.\ref{Phi}. We fitted these magnetization distributions using the function: $H(m,L)= B(L,T_c(L)) |m|^{2+1/\beta}$, with the value of the critical exponent for the magnetization: $\beta=0.3269$, being taken from Ref.\cite{tb96}. For small clusters $L=8,16$ the data deviates from the fit indicating a smaller $\beta$. For larger clusters the fit is more optimal. The tops of the distribution functions in Fig.\ref{Phi} are not entirely flat, but rather having small bumps. The integral weight of these bumps is small and decreases from $L=32$ to $L=64$.

Having found the dependence of the zero-mode Hamiltonian on the magnetization, the finite-size scaling determines the rest of it:
\begin{equation}\label{ZeroMode}
H[m]= L^{d-(\gamma-1)/\nu} \left( B(T) |m|^{2+1/\beta} + A(T) m^2 \right) -L^d h m
\end{equation}
where $A(T_c(L))=0$ and $\gamma$ is the critical exponent of the susceptibility. The fit to the width of the magnetization distributions in Fig.\ref{Phi} give us the five coefficients: $L^{(1+2\beta)/\nu} B(T_c(L))$, all accounted for by the critical exponent: $\nu=0.629(4)$. Our numerical algorithm gives us the five points of the bifurcation line. In the notations of inverse temperature: $K=J/T$, the Eq.(\ref{BifLine}) reads as $K_c(L)=K_c-k_c/L^{1/\nu}$. Using the above found critical exponent $\nu$ we establish the bifurcation line: $K_c=0.221657(8)$ and $k_c=0.213(4)$.

It is convenient to rescale the magnetization and the deviation of the temperature from the bifurcation line as
\begin{equation}
\mu= L^{\beta/\nu} m , \quad\quad \theta = L^{1/\nu} \left(T-T_c(L) \right) .
\end{equation}
Now, using the Josephson's hyperscaling relationship the zero-mode Hamiltonian simplifies:
\begin{equation}\label{ZeroModeIsing}
H[\mu]=-L^{d-\beta/\nu} h \mu +A(\theta) \mu^2+ B(\theta) |\mu|^{2+1/\beta} ,
\end{equation}
where $A(\theta)$ and $B(\theta)$ are analytic functions and $A(\theta_c)=0$. Although we have verified this Hamiltonian for the three-dimensional Ising model, the general derivation implies that it is correct for other phase transitions of the second order like in higher dimensions $d\ge 5$. In this case the zero-mode Hamiltonian was found by the microscopic calculations integrating out all inhomogeneous modes $\sigma_\mathbf{x}$ inside the cluster \cite{bz85}. The result is the Landau potential given by Eq.(\ref{ZeroMode}) for the mean-field critical exponents: $\beta=1/2$ and $\gamma=1$,
\begin{equation}
H[m]= L^{d}\left(-h m + B(\theta) m^4 +A(\theta) m^2 \right) .
\end{equation}
For the two-dimensional Ising model $\beta=1/8$ and we predict the analytic zero-mode Hamiltonian:
\begin{equation}
H[\mu]= -L^{15/8} h \mu + B(\theta) \mu^{10} +A(\theta) \mu ^2 .
\end{equation}
For the multicomponent $O(N)$ ferromagnets the zero-mode Hamiltonian is generalized as:
\begin{equation}
H[\mathbf{m}]= -L^{d-\beta/\nu} \mathbf{h}\cdot \mathbf{m} + B(\theta) (\mathbf{m}\cdot \mathbf{m})^{1+1/2\beta} +A(\theta) \mathbf{m}^2 ,
\end{equation}
where $\mathbf{m}$ is the vector order parameter. The limits of $\theta$ in these Hamiltonians is of the order of one. 

Given the explicit zero-mode Hamiltonian, many problems can be solved directly. For instance, the zero-mode part of the specific heat reaches maximum deep into the "ordered phase" at $\theta<0$. We also solve a problem of the critical cluster in external uniform field. At weak fields a linear Curie-law response takes place: $m\propto L^{2-\eta} h$, with the entire cluster responding as a whole. As the field grows much larger than a crossover field: $h^*= 1/L^{\delta\beta/\nu}$, the average order parameter can be found using the steepest descent approximation:
\begin{equation}
\langle m \rangle = \left(L^{(\gamma-1)/\nu} h/B_c \right)^{\beta/(1+\beta)} .
\end{equation}
This solution represents an enhancement of the response in the yellow domain in Fig.\ref{PhaseDiagram} centered on the bifurcation line. If the field $h$ increases simultaneously with the size $L$ then a new critical exponent $\beta/(1+\beta)$ holds.

The precise location of the bifurcation line is conveniently determined by the Binder cumulant: $U=\langle m^4\rangle/\langle m^2\rangle^2$ \cite{b81}. Its value on the bifurcation line:
\begin{equation}\label{Cumulant}
U(\theta_c)=\Gamma\left(\frac{\beta}{1+2\beta}\right)\Gamma\left(\frac{5\beta}{1+2\beta}\right)/ \Gamma^2\left(\frac{3\beta}{1+2\beta}\right), 
\end{equation}
is universal and is determined by the critical exponent $\beta$.

In conclusion, the effective zero-mode Hamiltonian for finite-size clusters of critical system undergoing a phase transition of the second order is found. Its dependence on the order parameter is asymptotically exact when approaching the bifurcation line and the critical point. In a certain sense, the energy term: $B |m|^{1+\Delta}$, describes a new critical phenomenon with the critical exponent being found from a new scaling relationship: $\beta\Delta= \beta+1$. The explicit zero-mode Hamiltonian with the Zeeman interaction allows one to consider, for instance, a system of critical magnetic dots coupled by the dipolar forces. 

We also  have devised an adaptive numerical method to determine the universality class of a critical system. One makes a guess about the critical exponent $\beta$ and then uses the criteria Eq.(\ref{Cumulant}) to drive the temperature towards the bifurcation line by progressively smaller steps while performing progressively longer runs of Metropolis numerical simulations, thus equilibrating the system. In the end, fitting the distribution of the order parameter one adjusts the value of the critical exponent $\beta$ and repeat the algorithm until it is converged. Once the critical exponent $\beta$ is known one scales down the finite-size $L$ to establish the critical exponent $\nu$.

I am grateful to V. Shadura for discussions and help with this paper.

\end{document}